\documentclass[a4paper,11pt]{article}

\usepackage[utf8]{inputenc}
\usepackage[T1,T2A]{fontenc}
\usepackage{amsmath,amsfonts,amssymb}

 \usepackage{graphicx}
\usepackage{wrapfig}
\usepackage{subfig}
\usepackage{hyperref}
\usepackage[dvipsnames]{xcolor}
\usepackage{dsfont}
\usepackage{amsmath,amssymb}

\begin{document}

\begin{center}
{\LARGE {\bf   Neutrino mixing and gravitational production via inflation }}
\end{center}

\vspace{0.3cm}
\begin{center}
{\bf Oleg Lebedev$^{\; 1}$ and Jong-Hyun Yoon$^{\; 2 }$}
\\ \ \\
$^{1}$  \it{Department of Physics and Helsinki Institute of Physics,\\
  Gustaf H\"allstr\"omin katu 2a, FI-00014 Helsinki, Finland}\\
  $^{2}$  \it{Department of Physics and Institute of Quantum Systems \\
  Chungnam National University, Daejeon 34134, Republic of Korea}\\
\end{center}

\vspace{0.7cm}
\noindent
{\bf Abstract:} 
We develop the Bogolyubov coefficient formalism for gravitational production of fermions with time-dependent mixing, which allows us to study a prototype neutrino system.
The neutrino masses and mixings depend on the scalar field values, i.e. the Higgs or singlet scalar expectation values. These are time-dependent in the Early Universe and, due to de Sitter fluctuations,
can reach very large values during inflation. 
As a result, 
gravitational production of  all types of  neutrinos can be much enhanced. 
We obtain an upper bound on the abundance of active and sterile  neutrinos produced by classical gravity, $Y \lesssim 10^{-11}$.

 \newpage

\tableofcontents

 \vspace{0.7cm}

\section{Introduction}

Gravitational particle production is an important aspect of the Early Universe cosmology  \cite{Parker:1969au}-\cite{Grib:1976pw}.
It is particularly efficient shortly after inflation  \cite{Starobinsky:1980te}-\cite{Linde:1981mu},   generating  an 
 irreducible background for model studies. Most importantly, it contributes to dark matter (DM) and dark relic production.
 The gravitational contribution is to be added to all other production sources, which makes non-thermal  DM model building highly non-trivial \cite{Lebedev:2022cic}.

Particle production via classical gravity is determined primarily by the particle effective mass as well as its non-minimal coupling to gravity. In realistic settings, the effective particle mass 
is an environment-dependent quantity. For example, the fermion mass is determined by the Yukawa coupling to a scalar, whose value evolves during and after inflation, unless
this scalar is super-heavy. Therefore, to evaluate gravitational particle production, one needs to take into account time dependence of the effective mass \cite{Feiteira:2025phi}, while in multi-field systems one should also
account for time-dependent particle mixing. The latter applies to the neutrino sector, in particular, and is the focus of our work.

During inflation with Hubble rate $H$, a minimally coupled real scalar field $s$ approaches the value
 \cite{Starobinsky:1986fx,Starobinsky:1994bd}
 \begin{equation}
 \langle s^2 \rangle \rightarrow {3H^4 \over 8\pi^2 m_s^2} ~~{\rm or}~~\sqrt{3\over 2\pi^2 }\, {\Gamma (3/4) \over \Gamma (1/4) }\, {H^2 \over \sqrt{\lambda_s}} \;,
 \label{variance}
 \end{equation}
 depending on which term dominates the scalar potential. Here,
$m_s\ll H$ is the scalar mass and $\lambda_s$ is the self-coupling, $V(s) = {1\over 2} m_s^2 s^2 + {1\over 4} \lambda_s s^4 \, .$
If inflation is not long enough for the scalar to reach the equilibrium value, the size of the scalar condensate is determined by the initial conditions  and its minimal value after 60 $e$-folds
is of order $H$  \cite{Feiteira:2025rpe}. In either case, the average scalar field value is very large. This makes the fermions with Yukawa couplings   to the scalar heavy. 

 During inflation, the fermions within each Hubble patch gain the same mass determined by the local value of $s(x)$, which is approximately constant inside the patch. Its average value over all the patches is given by the above variance (\ref{variance}).
Note that only the absolute value of $s$ matters due to the chiral symmetry in the fermion sector. The end result is that, on the average, the fermions are heavy and their mass is given by $Y \sqrt{ \langle s^2 \rangle }$, with $Y$ being the Yukawa coupling. 
After inflation, $\langle s^2 \rangle$ remains frozen for some time and subsequently decays, relaxing eventually to its current value. This introduces non-trivial time dependence in the fermion mass.

Gravitational fermion production in realistic settings with a time-dependent mass has been studied in Ref.\,\cite{Feiteira:2025phi}.   However, these results  are not  applicable to fermions with time-dependent mixing, including the neutrino sector.
In this case, the standard Bogolyubov coefficient formalism \cite{Bogolyubov:1958se}-\cite{Kolb:2023ydq}  fails due to  probability non-conservation in a single-fermion sector. 
Therefore, a generalization of this  approach is necessary. This is the subject of our Section \;\ref{sec2}.

We apply the generalized Bogolyubov approach to the prototype neutrino sector with a seesaw-like mass matrix. In the case of hierarchical mass matrix entries, the produced particle number can be evaluated analytically, while the other cases are treated
numerically. These  results allow us to assess the gravitational production efficiency of active and sterile neutrinos, with the latter being potentially relevant to the problem of dark matter \cite{Dodelson:1993je}-\cite{Cline:2026cek}.

Other aspects of gravitational fermion production have been explored in \cite{Shaposhnikov:2020aen}-\cite{Bertuzzo:2026gkj}. Scalar production and its subtleties have recently been discussed in  \cite{Feiteira:2025rpe}.
Gravitational vector production has its own peculiar features that have been studied in \cite{Graham:2015rva}-\cite{Lebedev:2025snd}.

\section{Gravitational fermion production: single species}
\label{single}

Gravitational production of a single fermion species in an expanding background has been studied by Parker in \cite{Parker:1971pt} and further developed in \cite{Chung:2011ck,Koutroulis:2023fgp,Feiteira:2025phi}.
 In what follows, we repeat this analysis and elucidate certain points which are important for its generalization to the multi-fermion case with mixing.

The Dirac equation in curved space has the form
\begin{equation}
(i \gamma^\alpha \nabla_\alpha- m) \,\Psi =0 \;.
\end{equation}
It is derived  from the action $\int d^4 x \sqrt{|g|} \bar \Psi (i \gamma^\alpha \nabla_\alpha- m) \,\Psi $, where $g_{\mu\nu}$ is the space-time metric, $\nabla$ is the covariant derivative and $\alpha$ is the  local Lorentz index.
The Friedmann metric in terms of the conformal time $x_0 \equiv \eta$ is  
\begin{equation}
ds^2 = a(x_0)^2 \, \eta_{\mu \nu } dx^\mu dx^\nu \;,
\end{equation}
with $\eta_{\mu \nu } $ being the Minkowski metric
The Weyl transformation
\begin{equation}
g_{\mu \nu }  = \Omega^2 \tilde g_{\mu \nu } ~,~ \Psi = \Omega^{-3/2} \tilde \Psi ~,~e_\alpha^\mu = \Omega^{-1} \tilde e_\alpha^\mu ~,
\end{equation}
where   $\Omega= a(x_0)$ and   $e_\alpha^\mu$ is the vierbein, 
  allows us to eliminate $a(x_0)$   from the action, apart from the mass term. 
Omitting  the tilde over the transformed quantities, the  Dirac equation now takes the form
\begin{equation}
\left[i \gamma^\mu \partial_\mu-  a(\eta) m\right] \,\Psi =0 \;.
\end{equation}
This is 
a flat space Dirac equation with a time-dependent mass. In general,  $a(\eta)$ and $m$   evolve in time, which leads to particle production. 
The scale factor evolves from $a(-\infty ) \rightarrow 0$ to $a(\infty ) \rightarrow \infty$, with the end of inflation corresponding to $\eta \sim 0$.

The solution is expressed in terms of the basis functions $\{U_i,V_i\}$ with constant coefficients,
\begin{equation}
 \Psi (x) = \sum_i  \left( a_i U_i + b_i^\dagger V_i \right) \;,
 \end{equation}
 where  $i$ denotes collectively the spin and momentum indices.
 In the Heisenberg picture, $a_i,b_i$ are the annihilation operators with 
  the standard  time-independent anti-commutation relations, $\{  a_i , a^\dagger_j\}= \delta_{ij}$, $\{  b_i , b^\dagger_j\}= \delta_{ij}$, etc.

The explicit form of the basis functions can be chosen as \cite{Chung:2011ck,Koutroulis:2023fgp}
\begin{eqnarray}
 &&U_{{\bf k} ,s} (\eta, {\bf x})= {e^{i {\bf k} \cdot {\bf x}}    \over (2 \pi)^{3/2}} 
 \left(
 \begin{matrix}
  u_{A,k} (\eta) \\
  s \, u_{B,k} (\eta)
 \end{matrix}
 \right)   \otimes h_s ({\bf \hat k}) \;,\\
 &&V_{{\bf k} ,s} (\eta, {\bf x})= -{e^{-i {\bf k} \cdot {\bf x}}    \over (2 \pi)^{3/2}} 
 \left(
 \begin{matrix}
  -u_{B,k}^* (\eta) \\
  s \, u_{A,k}^* (\eta)
 \end{matrix}
 \right) \otimes  h_s (-{\bf \hat k})  \, e^{i \phi}\;,
 \label{UV}
 \end{eqnarray}
where $k \equiv |{\bf k}|$, ${\bf \hat k} = {\bf k } / |{\bf k}| = (\theta, \phi)$ in spherical coordinates  and  $h_s$ are the helicity 2-spinors satisfying
\begin{equation}
 {\bf \hat k} \cdot \vec{ \sigma} \; h_s = s \, h_s ~~,~~ s= \pm 1 \;,
 \end{equation}
 with $\vec{ \sigma}$ being the Pauli matrices.
 The gamma matrices are taken to be of the form
 \begin{equation}
 \gamma^0 = 
 \left(
 \begin{matrix}
I & 0 \\
 0 & -I
 \end{matrix}
 \right) ~,~ 
 \gamma^i =  \left(
 \begin{matrix}
0& \sigma^i \\
 -\sigma^i & 0
 \end{matrix}
 \right) ~.
 \end{equation}
The 2-spinors satisfy
\begin{equation}
h^{\dagger}_s ({\bf \hat k} ) \,h_r ({\bf \hat k} ) = \delta_{rs} \;,
\end{equation} 
such that the basis orthonormality 
\begin{equation}
 (U_i, U_j)=  (V_i, V_j)    =\delta_{ij}~, ~  (U_i, V_j)= 0\;,
 \label{ortho}
 \end{equation}
requires the normalization condition
 \begin{equation}
|u_A|^2 + |u_B|^2 =1\;.
\label{norm}
 \end{equation}
The scalar product is defined as $(f,g) = \int d^3 x \, f^\dagger g$ 
 and  the spacial part of the wave functions is given  by the orthonormal set  $ {e^{i {\bf k} \cdot {\bf x}}    \over (2 \pi)^{3/2}} $.
 The explicit form of the spinors $h_s $ can be found in \cite{Chung:2011ck,Koutroulis:2023fgp}.

The Dirac equation with the above Ansatz reduces to 
\begin{equation}
 i\partial_\eta        
 \left(
 \begin{matrix}
  u_A \\
  u_B
 \end{matrix}
 \right)
  = \left(
 \begin{matrix}
am & k \\
 k & -am
 \end{matrix}
 \right) \; 
 \left(
 \begin{matrix}
  u_A \\
  u_B
 \end{matrix}
 \right)
 ~.~ 
 \label{EOM-u0}
 \end{equation}
This system can also be written as
\begin{eqnarray}
  && u_A^{\prime \prime} + \left [i(m a)^\prime +a^2 m^2 +k^2\right] \,u_A=0 \,,\\
  && u_B^{\prime \prime} + \left [-i(m a)^\prime +a^2 m^2 +k^2\right] \,u_B=0 \,,
  \label{uA-eq}
  \end{eqnarray}
  where the prime stands for the $\eta-$derivative.
The EOM 
 are to be solved with specific boundary conditions. The Bunch-Davies   initial condition   corresponds to the flat space vacuum $a_i | 0 \rangle = b_i | 0 \rangle =0$ \cite{Bunch:1978yq}, which 
defines the $in$ wavefunction:
at $\eta\rightarrow -\infty$, the $a(\eta) m$ terms vanish   and the positive frequency solution is given by 
\begin{equation}
 \left(
 \begin{matrix}
  u_A \\
  u_B
 \end{matrix}
 \right)^{\rm in}~ \overset{\small{\eta \rightarrow -\infty}}{\longrightarrow} 
 \left(
 \begin{matrix}
  {1/ \sqrt{2}} \\
 {1/ \sqrt{2}}
 \end{matrix}
 \right) \;e^{-i k \eta} \;.
 \end{equation}
 This condition fixes  $u_{A,B}$ uniquely.

Similarly, 
 at $\eta \rightarrow  \infty$, the system becomes  adiabatically flat and the evolution matrix  diagonal.  
 The positive eigenvalue solution corresponds to
  \begin{equation}
 \left(
 \begin{matrix}
  u_A \\
  u_B
 \end{matrix}
 \right)^{\rm out}~ \overset{\small{\eta \rightarrow \infty}}{\longrightarrow} 
 \left(
 \begin{matrix}
 1\\
0
 \end{matrix}
 \right) \;e^{-i \int \omega(\eta) d\eta} \;,
 \label{eta-inf}
 \end{equation}
with $\omega \rightarrow a(\eta) m$. 
The normalization of  $u_B$ is fixed by including the first order correction in $k/(am)$  to this asymptotic form. 
The resulting   eigenvector contains $k/(2am)$ instead of zero in the lower entry.

 The solutions with different boundary conditions are related to each other by a linear transformation with constant coefficients.  Under the basis change,
\begin{equation}
\tilde U_{{\bf k} ,s} = \alpha_{{\bf k} ,s} U_{{\bf k} ,s} + \beta_{{\bf k} ,s} V_{{\bf k} ,s} \;,
\end{equation}
where the tilded quantities refer to the  new basis. The basis orthonormality (\ref{ortho}) requires this transformation to be $unitary$.
Since $\Psi (x)$ remains intact,  the basis change necessitates a linear redefinition of the ladder operators.
Using the orthonormality condition, one finds that the Bogolyubov coefficient is given by 
\begin{equation}
\beta_{{\bf k} ,s} = {\rm phase } \times  (u_{A,k} \tilde u_{B,k} -u_{B,k} \tilde u_{A,k}) \;,
\label{Bog}
\end{equation}
where the constant   phase  is irrelevant for our purposes. 
The Bogolyubov coefficient is a measure of particle creation.
Identifying the tilded/un-tilded  objects  with $out$/$in$  quantities, one finds that 
$| \beta_{i}|^2 $ determines  the average particle number 
in the $in$ vacuum $|0\rangle$ with respect to the $out$ number operator,
\begin{equation}
 \langle \tilde N_{{\bf k} ,s} \rangle \equiv  \langle 0 | \tilde a_{{\bf k} ,s}^\dagger \tilde a_{{\bf k} ,s} | 0 \rangle = | \beta_{{{\bf k} ,s}}|^2  \;.
 \label{N}
\end{equation}
 The physical particle number density is  
 \begin{equation}
 n =\sum_s  \int {d^3 {\bf k} \over (2\pi)^3 a^3} \, |\beta_{{\bf k},s }|^2 \;,
 \label{n0}
 \end{equation}
 with the same number of anti-particles.
 The Bogolyubov coefficient (\ref{Bog}) is computed by determining the $in$ and $out$ solutions to the EOM. It is constant and can be calculated  at any point,
 \begin{equation}
\beta_{{\bf k} ,s}^\prime =0\;.
\end{equation}
The choice of the appropriate point depends on the specifics of the  setting.

\subsection{Comments and observations}

 The above exposition is quite standard, yet it is helpful to clarify a few points. 
 The convenience of the choice of the basis functions (\ref{UV})  is due to the relation
\begin{equation}
    -i\,s\, \sigma^2      \,  \left(
 \begin{matrix}
  u_{A}  \\
  s \, u_{B} 
 \end{matrix}
 \right)^*  =              \left(
 \begin{matrix}
  -u_{B}^*  \\
  s \, u_{A}^* 
 \end{matrix}
 \right)\;.      
 \end{equation}
Given a solution to (\ref{EOM-u0}), this relation allows one to construct another  linearly independent solution. This is due to the fact that the Pauli matrices $\sigma^1, \sigma^3$ anticommute with $\sigma^2$ and 
that $m$ is real.  If $(u_A,u_B)^T$ is an eigenstate of the Hamiltonian in (\ref{EOM-u0}) with a positive eigenvalue,  then  $(-u_{B}^* , 
    u_{A}^* )^{T}$ is an eigenstate with a negative eigenvalue.
Therefore, in the single fermion case, the form of the basis functions is dictated by simple properties of the Hamiltonian in  (\ref{EOM-u0}), which does not generalize to a multi-fermion system.
Any  solution to  (\ref{EOM-u0}) is a linear combination of the basis solutions, hence
\begin{equation}
  \left(
 \begin{matrix}
  u_{A}  \\
   \, u_{B} 
 \end{matrix}
 \right)^{\rm out}    = \alpha\;    \left(
 \begin{matrix}
  u_{A}  \\
   \, u_{B} 
 \end{matrix}
 \right)^{\rm in}       + \beta \;    \left(
 \begin{matrix}
  -u_{B}^*  \\
   \, u_{A}^* 
 \end{matrix}
 \right)^{\rm in}\;,   
 \label{lin}   
 \end{equation}
 where we suppress the indices for simplicity.
 Taking a scalar product of this relation with  $(-u_{B}^* , 
    u_{A}^* )^{\rm in \;T}$, one readily obtains the Bogolyubov coefficient  (\ref{Bog}).

It is also worth commenting on the spin structure of the fermion evolution.   The Dirac equation preserves  
  helicity,  i.e. the spin projection on the momentum direction,  even for a massive fermion.  This is due to the fact that the Dirac operator
  commutes with 
    $\Sigma^i \partial_i$, where
\begin{equation}
 \Sigma^i = 
\left(
\begin{matrix}
\sigma^i & 0 \\ 
 0 & \sigma^i
\end{matrix}
\right)\;.
\end{equation}
The commutator vanishes by virtue of 
 $\left[ \gamma^0, \Sigma^i \right]= 0 $  and $\epsilon^{ijk} \sigma_i \partial_j \partial_k =0$. 
In the expanding Universe, the spacial part of the  translational and rotational invariance remains intact, hence the corresponding 
quantum numbers can be used for classification of states. 
In particular, the momentum $\int d^3 x \Psi^\dagger (-i \nabla) \Psi$ is conserved.
The time translation symmetry is, on the other hand, broken and the different 
Hamiltonian eigenstates are mixed by the evolution. The Hamiltonian  $H = \int d^3 x \bar \Psi (-i \gamma \cdot \nabla + a(\eta)\,m) \Psi$ 
is time-dependent and the energy is not conserved.

The Dirac operator depends on the spacial momentum only through the helicity combination $-i\sigma^i \partial_i$, 
\begin{equation}
i \gamma^\mu \partial_\mu-  a(\eta) m\times \mathds{1}=  \left( 
\begin{matrix}
i\partial_0 -a\,m &  i\sigma^i\partial_i\\
 - i\sigma^i\partial_i &  -i\partial_0 -a\,m 
\end{matrix}
\right) \;,
\end{equation}
thus  $U_{{\bf k},s}$ and $V_{{\bf k},s}$ 
satisfy the same equation despite having different momentum signs. Hence, a general solution to the Dirac equation involves a linear combination of the two.

The wavefunction components $U_{{\bf k},1}$ and $V_{{\bf k},1}$ correspond to the positive eigenvalues  of the helicity operator $-i\Sigma^i \partial_i$,
as seen from (\ref{UV}) by relabelling ${\bf k'}\equiv - {\bf k}$.  A general Dirac equation solution with positive helicity is a linear combination of the two with constant coefficients.
This implies that the Bogolyubov transformation preserves helicity.
In the basis $(U_1, U_{-1}, V_1, V_{-1})^T$ for a given ${\bf k}$, it 
has a block diagonal form 
 \begin{equation}
  \left( 
\begin{matrix}
\alpha \times  \mathds{1} &  \beta \times  \mathds{1}\\
 -\beta^*  \times  \mathds{1} & \alpha^* \times  \mathds{1}
\end{matrix}
\right) \;.
\label{O}
\end{equation}

 These remarks are useful for generalizing the Bogolyubov formalism to multiple fermions with mixing. We note that the Bogolyubov transformation in flat space for fermions  with time-independent mixing
 has been considered before, e.g.  in the context of neutrino or neutron oscillations \cite{Blasone:2025atj,Tureanu:2018phm}. However, in realistic cosmological settings, both  fermion masses and mixings depend   
 on the environment and thus on time.

 \section{Bogolyubov formalism for fermions with time-dependent mixing}
 \label{sec2}

 Consider a system of 2 fermions $\Psi_1, \Psi_2$  which mix via the mass terms. 
 In  realistic settings, the fermion mass is determined by the value of some scalar field, which itself depends on the Universe expansion history. Therefore, the fermion mass is generally time-dependent.
 In what follows, 
we  focus on the seesaw-inspired     \cite{Minkowski:1977sc,Gell-Mann:1979vob,Yanagida:1979as,Mohapatra:1979ia}     mass matrix of the form\footnote{More realistic neutrino mass matrices are reviewed in \cite{Valle:2006vb}-\cite{SajjadAthar:2021prg}.} 
 \begin{equation}
 {\cal M} = \left(
\begin{matrix}
 0 & m (\eta) \\
m(\eta) & M
 \end{matrix}
 \right)\;,
 \label{m-matrix}
\end{equation}
where $m(\eta) $ is determined by the value of a  Higgs-like field. On the other hand, $M $ is taken to be approximately constant. This is possible when the corresponding scalar field is very heavy, with the mass 
exceeding the Hubble rate. We also assume that the zero entry in the mass matrix is enforced by symmetry as in the seesaw models. Both $m$ and $M$ are  taken to be real and positive using phase redefinition. In the Early Universe, $m(\eta)$ 
can be so large as to exceed  $M$, implying $maximal$ fermion mixing.

The Dirac equation becomes
\begin{equation}
\left[   \mathds{1} \times   i \not\skew{-6}\partial- a(\eta)\,  {\cal M}(\eta)\,
\right] \; \left(
\begin{matrix}
  \Psi_1 \\
  \Psi_2
 \end{matrix}
\right)=0\;.
\label{dirac-Phi}
\end{equation}
In the component form,
\begin{eqnarray}
&& i \not\skew{-6}\partial \Psi_1 -a(\eta)\,m(\eta)\,\Psi_2=0 \;, \\
&& i \not\skew{-6}\partial \Psi_2 -a(\eta)\,m(\eta)\,\Psi_1 -a(\eta)\,M\, \Psi_2=0\;.
\label{EOM-comp}
\end{eqnarray}

It is convenient to introduce 
\begin{equation}
\Phi \equiv \left( 
\begin{matrix}
\Psi_1 \\
\Psi_2
\end{matrix}
\right)\,,
\end{equation}
which  satisfies the EOM  (\ref{dirac-Phi}). 
Its mode expansion can be put in the form 
\begin{equation}
   \Phi (x) = \sum_{ s}  \int {d^3 {\bf k} \over (2 \pi)^{3/2}} \,\Biggl[ {e^{i {\bf k} \cdot {\bf x}}   } 
 \left(a^1_{{\bf k} s} \, {{\bf u}^1_{ {\bf k} s}} + a^2_{{\bf k} s}     \, {{\bf u}^2_{ {\bf k} s}}              \right)   \otimes h_s ({\bf \hat k}) 
\, +\,  {e^{-i {\bf k} \cdot {\bf x}}    } 
     \left(b^{1\,\dagger}_{{\bf k} s} \, {{\bf v}^1_{ {\bf k} s}} + b^{2\,\dagger}_{{\bf k} s}     \, {{\bf v}^2_{ {\bf k} s}     }         \right)   \otimes h_s (-{\bf \hat k})   \Biggr],
\nonumber
\end{equation}
where the normalization is chosen such that $\{a_{{\bf k}s}^i, a_{{\bf k'}s'}^{j\dagger}\} = \delta ({\bf k}-{\bf k'}) \, \delta_{ss'}\delta^{ij}$, etc. The modes 
${{\bf u}^1_{ {\bf k} s}} , {{\bf u}^2_{ {\bf k} s},  {{\bf v}^1_{ {\bf k} s}}, {{\bf v}^2_{ {\bf k} s}} } $ 
are the 4-vectors which depend on time $\eta$ and 
solve the EOM  for  fixed $k,s$:
\begin{equation}
i\partial_\eta \,{\bf f} = \left(
\begin{matrix}
0 & ks & am &0 \\ 
ks &  0 & 0 &-am \\
am & 0 & aM  & ks \\
0 & -am &ks & -aM
\end{matrix}
\right) \, 
{\bf f}  =H\, {\bf f}  \;,
\label{EOM-matrix}
\end{equation}
with $s=\pm 1$.
The $4\times 4$ evolution matrix  is recognized as the time-dependent Hamiltonian in the $\Psi^\dagger, \Psi$ basis. It is Hermitian, traceless and has a positive determinant,
which ensures that it possesses
2 positive and 2 negative eigenvalues. 

The ${\bf u}^i$ and ${\bf v}^i$ vectors are solutions to the EOM with specific boundary conditions.  
They are not, in general, eigenvectors of the time-dependent Hamiltonian $H$. Indeed, the Hamiltonian can be diagonalized by a time-dependent unitary transformation $R$: $R^\dagger HR = {\rm diag} (\lambda_1,..,\lambda_4)$,
which yields an equation for $\tilde {\bf f} \equiv R^\dagger {\bf f}$: $ i (R^\dagger \partial_\eta R) \tilde {\bf f} + i\partial_\eta \tilde {\bf f}  = {\rm diag} (\lambda_1,..,\lambda_4) \tilde {\bf f} $.  Due to the non-diagonal structure of
$R^\dagger \partial_\eta R$, the Hamiltonian eigenstates do not solve this equation.  Instead, ${\bf u}^i$ and ${\bf v}^i$  are chosen to be eigenstates of the Hamiltonian in the asymptotic regions    $\eta \rightarrow \pm \infty$,  where $H$ changes slowly 
and the concept of a particle is well  defined.
 The  ${\bf u}$'s and  ${\bf v}$'s are  split    according to the sign of the corresponding  eigenvalue $\lambda_H$ in a given asymptotic regime:
\begin{equation}
{\bf u}^i: \lambda_H >0 ~~,~~ {\bf v}^i: \lambda_H <0~.
\end{equation}
They are related by ``charge conjugation'', 
\begin{equation}
    {{\bf v}^{i}_{ {\bf k}s}}           =     {\;\rm diag\,}(i\sigma_2,i\sigma_2) \,{{\bf u}^{i*}_{ {\bf k}s}} \;,
    \label{charge-sym}
\end{equation}
which preserves the EOM while flipping the sign of the Hamiltonian eigenvalue.  
Any solution of (\ref{EOM-matrix}) is a linear combination of 4 independent basis solutions  ${\bf u}^i$, ${\bf v}^i$. 
However, the choice of the basis is not unique: one may set the boundary conditions at $\eta \rightarrow -\infty $ or $\eta \rightarrow +\infty $,
which results in different sets of ${\bf u}^i$, ${\bf v}^i$. Indeed, since they are not Hamiltonian eigenstates at all times, 
positive $H$ eigenstates at $\eta \rightarrow -\infty$ evolve into a mixture of positive and negative $H$ eigenstates at $\eta \rightarrow \infty$.
This is the essence of the Bogolyubov transformation.

The helicity flip is achieved by multiplying the solution with diag$(1,-1,1,-1)$, i.e.
\begin{equation}
{{\bf u}^i_{ {\bf k}, -1}} = {\rm diag} (1,-1,1,-1) \, {{\bf u}^i_{ {\bf k}, 1}} ~~,~~ {{\bf v}^i_{ {\bf k}, -1}} = {\rm diag} (1,-1,1,-1) \, {{\bf v}^i_{ {\bf k}, 1}}~.
\label{hel-sym}
\end{equation}
This allows us to focus on a single helicity state, say $s=1$, and double the states at the end of the calculation.

 Suppressing the momentum and spin indices, the anti-commutation relations and the orthonormality conditions  are {\it time-independent} and read
 \begin{eqnarray}
 &&  \{ a^i , a^{j\dagger} \}   =   \delta^{ij}~~,~~  \{ b^i , b^{j\dagger} \}   =   \delta^{ij} ~~,~~  \{ a^i, b^j \}   =  \{ a^i , a^j \} =\{ b^i, b^j \}  = 0 \;,
 \label{anti-comm} \\
 &&     ({\bf u}^i ,  {\bf u}^j)= \delta^{ij} ~~,~~   ({\bf v}^i ,  {\bf v}^j)= \delta^{ij} ~~,~~ ({\bf u}^i ,  {\bf v}^j)= 0\;, 
  \label{orth-con}
 \end{eqnarray}
 where $({\bf g}, {\bf h}) \equiv {\bf g}^\dagger {\bf h} $.
 This is ensured by the fact that ${\bf u}^i$, ${\bf v}^i$ are the Hamiltonian eigenstates in the asymptotic regions and 
 \begin{equation}
 \partial_\eta ({\bf g}, {\bf h}) =0 \;,
 \label{time-independence}
 \end{equation}
 for ${\bf g}, {\bf h}$ obeying  the EOM.
 
 The Bogolyubov transformation corresponds to a basis change in the space  $\{ {\bf u}_i ,{\bf v}_j \}$, which preserves the scalar products  (\ref{orth-con})   and thus is unitary.
 For explicit calculations, we need to choose a specific ordering of the vector entries. Let us rewrite the field operator as
 \begin{equation}
 \Phi = \sum_\alpha \left(a^1 {\bf U}^1 +      b^{1\dagger }   {\bf V}^1 +         a^2 {\bf U}^2      + b^{2\dagger }   {\bf V}^2   \right)_\alpha  \;,
 \end{equation}
 where $\alpha$ denotes collectively the 3-momentum and helicity; ${\bf U}^i, {\bf V}^j$ are the analogs of $U,V$ in the single-fermion case.
 Defining
\begin{equation}
 {\cal {U}} =   \left(
\begin{matrix}
 {\bf U}^1   \\
     {\bf V}^1  \\
    {\bf U}^2\\
    {\bf V}^2 
     \end{matrix}
\right)_\alpha~~,~~
{\bf { a}}= \left(
\begin{matrix}
a^1   \\
b^{1\dagger}  \\
  a^2\\
    b^{2\dagger}  
     \end{matrix}
\right)_\alpha \;,
\end{equation}
for a fixed $\alpha$, the field operator takes the form $  \Phi =\sum_\alpha  \left(   {\bf { a}}^T  {\cal {U}}  \right)_\alpha$.
Then, the Bogolyubov transformation that changes the basis while keeping $  \Phi $ intact corresponds to a unitary ${\cal O}$,
\begin{equation}
{\cal U}  \rightarrow     \tilde  {\cal U}  = {\cal O}   {\cal U} ~~, ~~ {\bf a}  \rightarrow \tilde {\bf a}   ={\cal O}^*   {\bf a}\;,
\end{equation}
 with ${\cal O}  {\cal O}^\dagger = 1$.
 The unitarity of ${\cal O}$ is also enforced by preservation of  the anti-commutation relations (\ref{anti-comm}).  Note that, for a fixed $\alpha$, 
 the functions ${\bf U}^i, {\bf V}^j$ satisfy the same orthonormality relations    (\ref{orth-con})   as ${\bf u}^i, {\bf v}^j$ do, up to
an overall factor $1/(2\pi)^3$.

 The total particle    number operator in the new basis is given by
 \begin{equation}
\tilde {\cal N}_{\rm tot} = \sum_{ \alpha} \left(     \tilde a^{1\dagger} \tilde a^1 +  \tilde a^{2\dagger} \tilde a^2 + \tilde b^{1\dagger} \tilde b^1  + \tilde b^{2\dagger} \tilde b^2    \right)_{\alpha} \;.
\end{equation}
The average number of ``tilded'' particles in the ``un-tilded''  vacuum defined by $a^i |0\rangle = b^i |0\rangle$ is $\langle 0|  \tilde {\cal N}  |0\rangle $. 
 For a fixed $\alpha$, the individual particle numbers are
 \begin{eqnarray}
 &&\langle     \tilde a^{1{\rm }\,\dagger} \, \tilde a^{1\, {\rm  }\,}   \rangle = \sum_{i=2,4}    \left|  {\cal O}^{1i}   \right|^2  = \sum_{i=1,2}           \left|  (\tilde {\bf u}_1, {\bf v}_i)      \right|^2~,~~
 \langle     \tilde a^{2{\rm }\,\dagger} \, \tilde a^{2\, {\rm  }\,}   \rangle = \sum_{i=2,4}    \left|  {\cal O}^{3i}   \right|^2  = \sum_{i=1,2}           \left|  (\tilde {\bf u}_2, {\bf v}_i)      \right|^2 , \nonumber \\
&&    \langle     \tilde b^{1{\rm }\,\dagger} \, \tilde b^{1\, {\rm  }\,}   \rangle = \sum_{i=1,3}    \left|  {\cal O}^{2i}   \right|^2  =    \sum_{i=1,2}           \left|  (\tilde {\bf v}_1, {\bf u}_i)      \right|^2     ~,~~
   \langle     \tilde b^{2{\rm }\,\dagger} \, \tilde b^{2\, {\rm  }\,}   \rangle = \sum_{i=1,3}    \left|  {\cal O}^{4i}   \right|^2  =\sum_{i=1,2}           \left|  (\tilde {\bf v}_2, {\bf u}_i)      \right|^2.
 \end{eqnarray}
As a result,
the charge conjugation (\ref{charge-sym}) symmetry ensures equal numbers of particles and anti-particles of each kind, while the helicity flip symmetry (\ref{hel-sym}) requires the same number of 
$s=1$ and $s=-1$ states,
 \begin{equation}
 \langle     \tilde a^{i{\rm }\,\dagger} \, \tilde a^{i\, {\rm  }\,}   \rangle = \langle     \tilde b^{i{\rm }\,\dagger} \, \tilde b^{i\, {\rm  }\,}   \rangle ~~,~~ \langle     \tilde a^{i{\rm }\,\dagger} \, \tilde a^{i\, {\rm  }\,}   \rangle_{s=1}=
  \langle     \tilde a^{i{\rm }\,\dagger} \, \tilde a^{i\, {\rm  }\,}   \rangle_{s=-1}\;.
 \end{equation}
 
 \subsection{$in$ and $out$ states}
 
 To compute the mode functions, we need to specify the boundary conditions. In the asymptotic regimes, curvature of the space-time plays no role and  the vacuum is the same as that in Minkowski space, i.e.
 the Bunch-Davies vacuum.  The $in$ states correspond to the flat space mode functions in the infinite past, while the $out$ states correspond to those in the infinite future.

  At $\eta \rightarrow -\infty$, the scale factor approaches zero and 
 \begin{equation}
H \rightarrow  \left(
\begin{matrix}
0 & ks & 0 &0 \\ 
ks &  0 & 0 &0\\
0& 0 & 0 & ks \\
0 & 0&ks & 0
\end{matrix}
\right) \, .
 \label{Hpast}
\end{equation}
 The  corresponding $in$  mode functions are the Hamiltonian eigenstates of the single-species type,
 \begin{equation}
{{\bf u}^1_{\rm in},{\bf v}^1_{\rm in}} (-\infty)=   {e^{\mp ik\eta} \over \sqrt{2} }\left(
\begin{matrix}
1 \\
  \pm 1\\
 0\\
 0  
 \end{matrix}
\right) ~~,~~
  {{\bf u}^2_{\rm in},{\bf v}^2_{\rm in}} (-\infty)=   {e^{\mp ik\eta} \over \sqrt{2} }\left(
\begin{matrix}
0\\
  0\\
1 \\
 \pm 1  
 \end{matrix}
\right) \;,
\label{initial-in}
\end{equation}
where the upper and lower signs apply to the ${\bf u}$ and ${\bf v}$ states, respectively, and we  have set $s=+1$.
 
 At $\eta \rightarrow \infty$, the  mass parameters $m,M$ become constant,  while the scale factor grows indefinitely and
 \begin{equation}
H \rightarrow a(\eta)\, \left(
\begin{matrix}
0 & 0 & m_\infty &0 \\ 
0 &  0 & 0 &-m_\infty \\
m_\infty & 0 & M  & 0 \\
0 & -m_\infty &0 & -M
\end{matrix}
\right) \, ,
\end{equation}
where $m_\infty \equiv m(\infty)$.
 The matrix has a block form with each block proportional to $\sigma_3$, which can thus be factored out. Hence,  it is diagonalized with the help of a $2\times 2$ orthogonal matrix ${\bf O}$:
 \begin{equation}
 {\bf O}^T  \,  \left( 
\begin{matrix}
0 & m_\infty \\ 
m_\infty &  M 
\end{matrix}
\right)  \,  {\bf O} = {\rm diag} (m_1,m_2) ~~,~~
 {\bf O}= \left(
 \begin{matrix}
\cos \phi & \sin\phi \\ 
-\sin\phi &  \cos\phi
\end{matrix}
 \right)\;,
\end{equation}
where
\begin{equation}
   \tan 2\phi = {2m_\infty \over M}~~,~~ m_{1,2}= {M\over 2} \mp   \left(     {M\over 2} \, \cos 2\phi +m \sin 2\phi            \right) \;. 
\end{equation}
The Hamiltonian is diagonalized by the $4\times 4$ orthogonal transformation ${\bf \tilde O}= {\bf O}  \otimes \mathds{1}$,
 \begin{equation} 
 {\bf \tilde O}^T H  {\bf \tilde O}= a(\eta) \times {\rm diag} \,(m_1,-m_1,m_2,-m_2)  \;.
 \end{equation}
Note that $m_1 < 0$, $m_2 >0$ and $|m_1| \ll m_2$. Thus, according to our definition of the ${\bf u}$ and ${\bf v}$ states,  ${\bf u}^1$ is associated with the $-m_1$ eigenvalue. 
Let us mark the eigenvectors in the mass-diagonal basis by a tilde, then  
 the $out $  states  at $\eta \rightarrow \infty$ read 
 \begin{equation}
{\bf \tilde u}^1_{\rm out} =\left( 
\begin{matrix}
0\\
 1\\
0\\
0 
\end{matrix}
\right) \; e^{im_1 f(\eta)} ,~~ 
{\bf \tilde v}^1_{\rm out} =\left( 
\begin{matrix}
1\\
 0\\
0\\
0 
\end{matrix}
\right) \; e^{-im_1 f(\eta)} ,~~ 
{\bf \tilde u}^2_{\rm out}= \left( 
\begin{matrix}
0\\
 0\\
1\\
0 
\end{matrix}
\right) \; e^{-im_2 f(\eta)} ,~~ 
{\bf \tilde v}^2_{\rm out}= \left( 
\begin{matrix}
0\\
 0\\
0\\
1 
\end{matrix}
\right) \; e^{im_2 f(\eta)} ,
 \end{equation}
 where 
 $f(\eta) = \int^\eta a(\eta') \,d\eta'$. Here, we have omitted overall constant phases, which are irrelevant to our calculations. 
 Rotating back to the original basis, we get the following ${\bf v} $-states:
 \begin{equation}
{\bf v}_{\rm out}^1(\infty)=  \left( 
\begin{matrix}
\cos\phi   \\
\cos\phi  \;k/(2am_1)\\
-\sin\phi   \\
-\sin\phi  \;k/(2am_1)
\end{matrix}
\right) 
\; e^{-im_1 f(\eta)} ,~~ 
{\bf v}_{\rm out}^2 (\infty)=  \left( 
\begin{matrix}
-\sin\phi \; k/(2am_2) \\
\sin\phi  \\
-\cos\phi  \; k/(2am_2)\\
\cos\phi 
\end{matrix}
\right)\;e^{im_2 f(\eta)}  \;.
\label{v-out-exact}
\end{equation}
Here, we have included the first order $k/(a m_i)$ corrections which can be important for some applications.  
The ${\bf u} $-states are obtained by applying   charge conjugation to the above vectors. In our numerical analysis, we use the exact Hamiltonian eigenvectors at large $|\eta|$ and do not resort to the $k/(am)$ expansion. 

We are interested in production of the lighter eigenstate with mass $|m_1|$. The corresponding particle number can be computed via
\begin{eqnarray}
&&N_1 (k) =  
\left| ({\bf u}_{\rm in}^1,   {\bf v}_{\rm out}^1   ) \right|^2 + \left| ({\bf u}_{\rm in}^2,   {\bf v}_{\rm out}^1   ) \right|^2 \;,
\label{N1-result}
\end{eqnarray}
for a fixed $s$. 
This yields the number of $out$ particles in the $in$ vacuum.
Due to (\ref{time-independence}), the result is time-independent and $N_1 (k) $ can be computed at any convenient reference point $\eta$.

 \section{Production of (almost) massless neutrinos by gravity}
 
 In this section, we study gravitational production of mixed fermions as motivated by  the seesaw neutrinos. Our considerations apply, however,  more generally and, in particular, to the production of long-lived right-handed neutrinos
 with the mass matrix (\ref{m-matrix}). 
 
 Some of the entries of the seesaw neutrino mass matrix depend on the Higgs field value and thus are time dependent. In the Early Universe, the Higgs field can take on very large values, of order the Hubble rate or above,
 while at late times it relaxes to the electroweak scale. How exactly this relaxation occurs is a model dependent question. In what follows, we explore 2 simplest options \cite{Feiteira:2025phi}, which provide us with the benchmark values 
 for the efficiency of gravitational light neutrino  production.
 
  During inflation, the Higgs takes on the value\footnote{This assumes the minimally coupled Higgs field. For a significant positive non-minimal coupling to gravity or to the inflaton, the Higgs gains a large effective mass driving it to zero.} 
 \begin{equation}
\langle h^2 \rangle \rightarrow  0.1 {H_e^2 \over \sqrt{\lambda_h}} \;,
\end{equation}
 where $H_e$ is the inflationary Hubble rate. Immediately after inflation, it remains frozen until the Hubble rate decreases to the level of the effective Higgs mass, $H^2 \sim  \lambda_h \langle h^2 \rangle  $. After that, it starts oscillating and decaying \cite{Enqvist:2015sua}.
Thus, the main stages in the Higgs evolution are: 
 \begin{itemize}
   \item{during inflation, it takes on a value of order $H_e$}
   \item{after inflation, remains frozen  until the Hubble rate becomes comparable to  the effective Higgs mass}
   \item{oscillates in the quartic potential and decays into the SM states}
   \item{takes on the electroweak value  at late times}
   \end{itemize}

 The Higgs-induced mass terms follow a similar evolution. However, the precise dynamics  are not known since the fermion masses are affected by the collective phenomena, e.g. non-perturbative and thermal effects. Hence, in what follows,
 we will consider two simplified benchmark scenarios,  which exhibit a fast and  a slow mass term decrease. Presumably, reality lies somewhere in between. 
 
 We focus on the regime where the Dirac mass during inflation exceeds the Majorana mass, $m(\eta\sim 0) \gg M$, which makes the effect  of the mixing more pronounced. At late times, however, the Dirac term becomes small and the light neutrino masses
 are determined by the usual seesaw relation. For simplicity, we take $M$ to be time-independent, 
which is justified in models where $M$ is determined by a VEV of a super-heavy scalar with mass exceeding the Hubble rate.

 In our calculations, we assume that the Universe is dominated by radiation after inflation. This is the case  when the inflaton potential is locally quartic or when reheating takes place quickly.  
 The scale factor and the Hubble rate can be approximated by the following functions:
\begin{eqnarray}
 && a(\eta) = \left\{  
 \left(  {1\over a_e H_e}  -\eta  \right)^{-1} H_e^{-1}
  ~~{\rm  for }~~ \eta\leq 0 ~~,~~ a^2_e H_e \left( \eta + {1\over a_e H_e}\right) ~ ~{\rm for}~ \eta>0 \right\}~,\\
 && H(\eta) = \left\{H_e ~~{\rm for}~~ \eta \leq 0 ~~,~~ H_e \, (a_e/a)^2 ~ ~{\rm for}~ ~\eta>0 \right\}~,
 \end{eqnarray}
 Inflation ends around $\eta \sim 0$, when the scale factor is close to $a_e$, after which $a$       scales approximately as  $\eta$.

 \subsection{Sharp mass drop}

A sharp fermion mass decrease  after the end of inflation can be modeled by the $\theta$-function:
\begin{equation}
m (\eta) = m \, \theta (\eta_0 -\eta) + \mu \, \theta (\eta - \eta_0 )~,
\end{equation}
 with $\eta_0 >\eta_e$ and  $m \gg \mu$.
 We parametrize 
 \begin{equation}
\eta_0 = N\; \eta_e \simeq  N \; {1\over a_e H_e} \;,
\end{equation}
where $a_e$ and $\eta_e $ correspond to the end of inflation. In practice,  for a Higgs-induced fermion mass,  $N$ is of order a few, depending on the precise value of $\lambda_h (H_e)$, and we take $N=5$ as the benchmark value.

 The number of produced particles is obtained by solving (\ref{EOM-matrix}) with the $in$ and $out$ boundary conditions and using (\ref{N1-result}). This can be done numerically for a wide range of the parameters. In what follows, we estimate the 
 particle number using simple analytical approximations.
 
 Let us focus on the regime $\eta <\eta_0$ and 
 \begin{equation}
 m(\eta) \gg M \;,
 \end{equation}
 implying maximal fermion mixing in the Early Universe.
 Neglecting $M$, the Dirac equation for $\Psi_{1,2}$ takes the form
 \begin{eqnarray}
&& i \not\skew{-6}\partial \Psi_1 -am\,\Psi_2=0 \;, \\
&& i \not\skew{-6}\partial \Psi_2 -am\,\Psi_1  =0\;.
\label{EOM-comp-1}
\end{eqnarray} 
 Introducing
 \begin{equation}
  \Psi_-= \Psi_1 -\Psi_2 ~~,~~   \Psi_+= \Psi_1 +\Psi_2 \;,
  \end{equation}
we find that the EOM for these spinors decouple,
\begin{eqnarray}
&& i \not\skew{-6}\partial \Psi_+ -am\,\Psi_+=0 \;, \\
&& i \not\skew{-6}\partial \Psi_- +am\,\Psi_-  =0\;.
\label{EOM-comp-2}
\end{eqnarray}
These equations correspond to the single fermion cases and can be solved analytically. 
Since the Bogolyubov coefficients can be computed at any point, let us evaluate the wavefunctions at $\eta_0$.
In what follows, we first calculate the $in$-wavefunctions  $\Psi_+(\eta_0)$ and $\Psi_-(\eta_0)$, while the $out$ wavefunctions are considered afterwards.\\ \ \\
 \noindent
{\underline{{\bf$ \Psi_+$} $in$-solution}}
\\ \  \\ 
The EOM for $\Psi_+\,$   has the standard  form so we can use the usual single fermion  $U,V$ decomposition (\ref{UV}). The system reduces to 
 \begin{eqnarray}
  && u_{A_+}^{\prime \prime} + \left [i(m a)^\prime +a^2 m^2 +k^2\right] \,u_{A_+}=0 \,,\\
  && u_{B_+}^{\prime \prime} + \left [-i(m a)^\prime +a^2 m^2 +k^2\right] \,u_{B_+}=0 \,.
  \label{uA-eq}
  \end{eqnarray}
 The $in$ boundary condition corresponding to  both ${\bf u}^1_{\rm in}$ and ${\bf u}^2_{\rm in}$ vectors (\ref{initial-in}) is 
\begin{equation}
   \left(
 \begin{matrix}
  u_{A_+} \\
  u_{B_+}
 \end{matrix}
 \right)^{\rm in}_{-\infty} =
  \left(
 \begin{matrix}
  {1/ \sqrt{2}} \\
 {1/ \sqrt{2}}
 \end{matrix}
 \right) \;e^{-i k \eta} \;.
\end{equation}
This implies that the solution has the standard single-fermion form.
Therefore, 
 during inflation, we have
 \begin{eqnarray}
&& u_{A_+}^{\rm in} (a) =  \sqrt{\pi k  \over 4 a H_e}\, 
  e^{ i {\pi \over 2} (1-im/H_e)}\, H^{(1)}_{{1/ 2} -{im/H_e}} \left({k\over a H_e}\right) \;,
   \label{uBinf}\\
  && u_{B_+}^{\rm in } (a) = \sqrt{\pi k  \over 4 a H_e}\, 
  e^{ i {\pi \over 2} (1+im/H_e)}\, H^{(1)}_{{1/ 2} +{im/H_e}} \left({k\over a H_e}\right) \;.
  \label{uAinf}
 \end{eqnarray}
 We are interested in the superhorizon modes at the end of inflation, i.e. the momentum range  ${k\over a_e H_e} \ll 1$, which can potentially have significant occupation numbers.
 Hence, we may use  
  the small argument expansion
 \begin{equation}
 H_\nu^{(1)} (x) \simeq - {i 2^\nu \Gamma (\nu)    \over \pi } \; x^{-\nu} \;.
 \end{equation}
 At the end of inflation ($a \sim a_e$), we thus find 
 \begin{eqnarray}
 &&  u_{A_+}^{\rm in} (a) \simeq {1\over \sqrt{2}}     \times e^{i {m\over H_e} \, \ln {k\over a H_e} } \;, \nonumber \\
   &&  u_{B_+}^{\rm in} (a) \simeq {1\over \sqrt{2}}     \times e^{-i {m\over H_e} \, \ln {k\over a H_e} } \;.
   \label{uAin-final}
\end{eqnarray}
Here, we have only kept the leading terms of order $m/H_e \ll 1$, enhanced by the large logarithm  $\ln {k\over a H_e} $.

Let us now consider how the wavefunction evolves after inflation ends, during the period $\eta_e < \eta < \eta_0$.
To leading order in $m/H_e$, the equation for $u_{A_+}^{\rm in} $ reads
\begin{equation}
  u_{A_+}^{{\rm in\, }\prime \prime} +  \alpha\, u_{A_+}^{\rm in} =0 \;,
  \end{equation}
 with $$ \alpha \simeq k^2 + im a_e^2 H_e \;.$$
 Since the frequency is constant, the solution is
 \begin{equation}
 u_{A_+}^{\rm in} = a_1 \, e^{i\sqrt{\alpha} \eta} + a_2 \, e^{-i\sqrt{\alpha} \eta} \;,
 \label{a1a2}
 \end{equation}
 where the constants are determined by the boundary conditions at the end of inflation. A similar result applies to $u_B$ up to the sign flip of $m$.
The coefficients are fixed by the wavefunction value and its derivative at $\eta_e$,
\begin{eqnarray}
 && a_1 = {1\over 2} \left[   u_{A_+}^{\rm in}(\eta_e) -i { u_{A_+}^{\rm in\, \prime}(\eta_e) \over \sqrt{\alpha}}   \right] \,e^{-i\sqrt{\alpha} \,\eta_e} \;, \\
  && a_2 = {1\over 2} \left[   u_{A_+}^{\rm in}(\eta_e) +i {u_{A_+}^{\rm in\, \prime}(\eta_e)  \over \sqrt{\alpha}}   \right] \,e^{i\sqrt{\alpha} \,\eta_e} \;,
  \label{solutions}
  \end{eqnarray}
  where $u_{A_+}^{\rm in}(\eta)$ is given by (\ref{uAin-final}).
  
Consider the small momentum regime 
\begin{equation}
   k \ll k_* =\sqrt{a_e^2 mH_e}  \;.
\end{equation}
In this case, $\alpha \simeq ia_e^2 m H_e$. Next, we expand the exponentials 
$e^{i\sqrt{\alpha} (\eta-\eta_e)}$ in (\ref{a1a2}) using 
$   \sqrt{\alpha} (\eta-\eta_e) \ll 1   $
and   $N^2 \gg N~, ~ N-1 \simeq N$.  At the second order in 
  $N \sqrt{m/H_e}$, we find
\begin{eqnarray}
&&u_{A_+}^{\rm in} (\eta_0)\simeq u_{A_+}^{\rm in}(\eta_e) \times \left( 1- {i\over 2} \, N^2 \, {m\over H_e} \right) \;, \\
&&u_{B_+}^{\rm in} (\eta_0)\simeq u_{B_+}^{\rm in}(\eta_e) \times \left( 1+ {i\over 2} \, N^2 \, {m\over H_e} \right)\;.
\label{relations}
\end{eqnarray}
 Our final result  for the momentum range $k<k_*$ is therefore,
\begin{eqnarray}
&& u_{A_+}^{\rm in} (\eta_0)  \simeq    {1\over \sqrt{2}}   \, e^{i {m\over H_e} \, \ln {k\over a_e H_e} }  \times \left( 1- {i\over 2} \, N^2 \, {m\over H_e} \right) \;, \\
&& u_{B_+}^{\rm in} (\eta_0)  \simeq {1\over \sqrt{2}}     \, e^{-i {m\over H_e} \, \ln {k\over a_e H_e} }  \times \left( 1+ {i\over 2} \, N^2 \, {m\over H_e} \right) \;,
\end{eqnarray}
which applies to both ${\bf u}^1_{\rm in}$ and ${\bf u}^2_{\rm in}$ boundary conditions (\ref{initial-in}). 

For higher momenta, the frequency becomes dominated by the $k^2$ term in $\alpha$, which is universal for ${u}_A$ and ${ u}_B$. Thus, the wavefunction only receives an overall
phase in this case, which does not affect the Bogolyubov coefficient.
\\ \ \\
 \noindent
{\underline{{\bf$ \Psi_-$} $in$-solution}}
\\ \  \\ 
The EOM for 
$ \Psi_-$ is obtained from that for $ \Psi_+$ by 
\begin{equation}
 m \rightarrow -m \;.
 \end{equation}
This amounts to swapping 
$ u_A \leftrightarrow u_B\;.$
The   $in$ boundary conditions for $\Psi_-$  corresponding to   ${\bf u}^1_{\rm in}$ and ${\bf u}^2_{\rm in}$ vectors (\ref{initial-in}) now differ,
\begin{equation}
   \left(
 \begin{matrix}
  u_{A_-} \\
  u_{B_-}
 \end{matrix}
 \right)^{\rm in\,, 1}_{-\infty} =
  \left(
 \begin{matrix}
  {1/ \sqrt{2}} \\
 {1/ \sqrt{2}}
 \end{matrix}
 \right) \;e^{-i k \eta} ~~~,~~~
  \left(
 \begin{matrix}
  u_{A_-} \\
  u_{B_-}
 \end{matrix}
 \right)^{\rm in\,, 2}_{-\infty} = -
  \left(
 \begin{matrix}
  {1/ \sqrt{2}} \\
 {1/ \sqrt{2}}
 \end{matrix}
 \right) \;e^{-i k \eta} \;.
\end{equation}
 The inflationary solutions are are thus given by 
\begin{eqnarray}
&& u_{A_{-(1)}}^{\rm in} (a) =  \sqrt{\pi k  \over 4 a H_e}\, 
  e^{ i {\pi \over 2} (1+im/H_e)}\, H^{(1)}_{{1/ 2} +{im/H_e}} \left({k\over a H_e}\right) \;,
   \label{uBinf}\\
  && u_{B_{-(1)}}^{\rm in} (a) = \sqrt{\pi k  \over 4 a H_e}\, 
  e^{ i {\pi \over 2} (1-im/H_e)}\, H^{(1)}_{{1/ 2} -{im/H_e}} \left({k\over a H_e}\right) \;,
  \label{uAinf1} \\
&& u_{A_{-(2)}}^{\rm in} (a) =  -\sqrt{\pi k  \over 4 a H_e}\, 
  e^{ i {\pi \over 2} (1+im/H_e)}\, H^{(1)}_{{1/ 2} +{im/H_e}} \left({k\over a H_e}\right) \;,
   \label{uBinf}\\
  && u_{B_{-(2)}}^{\rm in} (a) = -\sqrt{\pi k  \over 4 a H_e}\, 
  e^{ i {\pi \over 2} (1-im/H_e)}\, H^{(1)}_{{1/ 2} -{im/H_e}} \left({k\over a H_e}\right) \;.
  \label{uAinf2}
 \end{eqnarray}
After inflation, the wavefunctions receive the factors analogous to those in (\ref{relations}) up to $m \rightarrow -m$. The result is 
\begin{eqnarray}
 && u_{A_{-(1)}}^{\rm in} (\eta_0)  \simeq {1\over \sqrt{2}}     \, e^{-i {m\over H_e} \, \ln {k\over a_e H_e} }  \times \left( 1+ {i\over 2} \, N^2 \, {m\over H_e} \right) \;, \\ 
&& u_{B_{-(1)}}^{\rm in}  (\eta_0)  \simeq    {1\over \sqrt{2}}   \, e^{i {m\over H_e} \, \ln {k\over a_e H_e} }  \times \left( 1- {i\over 2} \, N^2 \, {m\over H_e} \right) \;, \\
 && u_{A_{-(2)}}^{\rm in}  (\eta_0)  \simeq  -{1\over \sqrt{2}}     \, e^{-i {m\over H_e} \, \ln {k\over a_e H_e} }  \times \left( 1+ {i\over 2} \, N^2 \, {m\over H_e} \right) \;, \\ 
&& u_{B_{-(2)}}^{\rm in}  (\eta_0)  \simeq   - {1\over \sqrt{2}}   \, e^{i {m\over H_e} \, \ln {k\over a_e H_e} }  \times \left( 1- {i\over 2} \, N^2 \, {m\over H_e} \right) \;. 
\end{eqnarray}
\\ \ \\
 Now we may compute $\Psi_1 = {1\over 2} \, (\Psi_+ + \Psi_-)~,~ \Psi_2 = {1\over 2} \, (\Psi_+ - \Psi_-)$.
 The  leading  $  m/H_e$ expansion result is given by
\begin{equation}
{\bf u}_{\rm in}^1 (\eta_0)\simeq {1\over \sqrt{2}} \,\left(
\begin{matrix}
1  \\
  1\\
 i {m\over H_e} \, \gamma \\
 -  i {m\over H_e} \, \gamma
 \end{matrix}
\right)~~,~~
{\bf u}_{\rm in}^2 (\eta_0)\simeq    {1\over \sqrt{2}} \, \left(
\begin{matrix}
  i {m\over H_e} \, \gamma \\
 -  i {m\over H_e} \, \gamma \\
1 \\
1  
 \end{matrix}
\right)  ,
\label{app-eta0}
\end{equation}
where 
  \begin{equation}
 \gamma = -{N^2 \over 2} +    \ln {k\over a_e H_e}   \;.
 \end{equation}
 We observe that the main effect of the wavefunction evolution is that the initial zeros get filled with small imaginary entries suppressed by $m/H_e$.
 \\ \ \\
 \noindent
{\underline{{\bf $out$-solution}}
\\ \  \\ 
At late times, the lighter fermion becomes effectively massless since $m(\eta) \ll M $ for $\eta \rightarrow \infty$. Thus,
in our analytical estimates, it suffices to consider the massless limit $m\rightarrow 0$. In our numerical analysis, however, we use the exact boundary conditions.
Consider the approximation
\begin{equation}
\eta> \eta_0\,:~~ m(\eta) \simeq 0 \;.
\end{equation}
In this case,  $\Psi_1$ and $\Psi_2$ decouple, and the ``massless'' fermion $\Psi_1$ satisfies
\begin{equation}
i\gamma^\mu \partial_\mu  \Psi_1=0\;.
\end{equation}
 This is the single fermion EOM solved in Sec.\,\ref{single}.

 The resulting equation for the spinor components is
\begin{eqnarray}
  && u_{A}^{{\rm out}\,\prime \prime} +  k^2 \,u_{A}^{\rm out}=0 \,,\\
  && u_{B}^{{\rm out}\, \prime \prime} +  k^2 \,u_{B}^{\rm out}=0 \,.
  \label{u-out-eq}
  \end{eqnarray}
 We drop the subscript ``1'' since we focus exclusively on the lighter state.
 These equations are solved by the exponentials with a constant frequency $\omega =k$.
The corresponding Hamiltonian is 
\begin{equation}
   \left(
\begin{matrix}
0 & k\\
k & 0
\end{matrix}
\right)   \;.
\end{equation}
We are interested in the ${\bf v}_{\rm out}^1$ vector, which corresponds to the negative eigenvalue of this Hamiltonian. 
The $\Psi_2$ boundary value is set to zero such that ${\bf v}_{\rm out}^1$ describes a propagation eigenstate of the $\Psi_1$ field.
Therefore, the $out$ boundary condition at $\eta\gg \eta_0$ is
\begin{equation}
{\bf v}_{\rm out}^1 (\eta) =  {1\over \sqrt{2}}\; \left(
\begin{matrix}
1\\
-1\\
0\\
0
\end{matrix}
\right) \; e^{+ik\eta}\;.
\label{out-v-0}
\end{equation}
Since the frequency is constant,  this form is valid at all $\eta \geq \eta_0$.

 \subsubsection{Particle number}
 
 The wavefunction is continuous and the particle number can be computed at $\eta =\eta_0$ using (\ref{N1-result}).
 We observe that ${\bf v}_{\rm out}^1 $ and ${\bf u}_{\rm in}^1$ are orthogonal. Therefore, the massless mode is produced entirely due to the mixing between $\Psi_1$ and $\Psi_2$
 in the Hamiltonian, which induces non-zero upper components in ${\bf u}_{\rm in}^2$.
  For $k \ll k_* = \sqrt{a_e^2 m H_e}$, the particle number is 
  \begin{eqnarray}
&&N_1 (k) =   \left(       -{N^2\over 2}    + \ln {k\over a_e H_e}    \right)^2 \, {m^2 \over H_e^2}\;.
\end{eqnarray} 
Unless the momentum is very low, the result is dominated by the $N^2/2$ term. For $k\gg k_*$, $N_1(k)$ drops and can be neglected. We see that the average occupation number is small, but the
occupied momentum shells extend to very large values of order $k_*$, which makes the overall effect important.

Our numerical result of solving the full evolution equation (\ref{EOM-matrix}) with a 4$\times$4 Hamiltonian is presented in Fig.\,\ref{N1-step}. We find good agreement with the analytical estimates both in the magnitude of $N_1 (k)$ and the position of the cutoff $k_*$.
We have verified that a small $\mu \ll M, H_e$ has essentially no effect on the results. In this case, the $out$ boundary conditions are of the type (\ref{v-out-exact}) and the evolution from $\eta=\infty $ to $\eta_0$ is non-trivial. 
Nevertheless, on dimensional grounds,  the correction to the massless result scales as $\mu/M,\, \mu/H_e$ and can be neglected, which we have confirmed numerically.

 \begin{figure}[h!]
    \centering
    \includegraphics[width=0.49\textwidth]{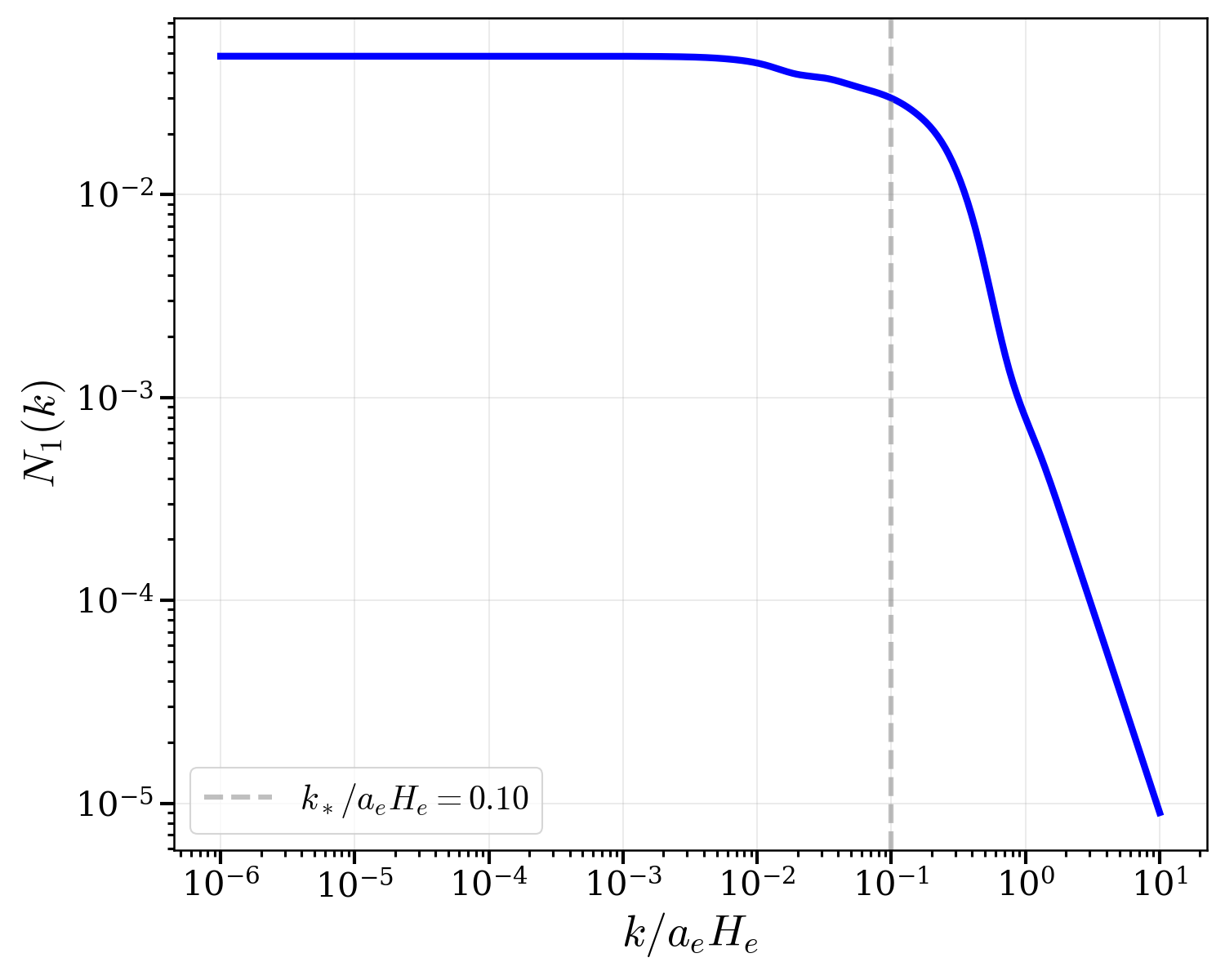}
    \caption{ The average occupation  number $N_1(k)$ for a step-function $m(\eta)$, with
    $m/H_e = 10^{-2}, M/H_e = 10^{-5}, N = 5, \mu \simeq 0 $. The dashed line marks the cutoff $k_*= a_e\,\sqrt{mH_e}$, beyond which particle production drops. 
    } 
    \label{N1-step}
\end{figure}

The total light fermion number density is 
  \begin{equation}
 n_1 \simeq \sum_{\rm d.o.f.}  \int {d^3 {\bf k} \over (2\pi)^3 a^3} \, N_1 (k) \, \theta (k_*-k)\;,
 \label{n0}
 \end{equation}
 where the sum includes the spin states and anti-particles.
 The integral can be computed analytically with the result
 \begin{equation}
 n_1 \simeq {\# {\rm d.o.f.}} \times {m^{7/2} \over 24\pi^2 \,H_e^{1/2}}  \, \left(       {N^2}    +\ln {H_e\over  m}    \right)^2\, \left( {a_e\over a}\right)^3 \;.
  \label{n1-tot1}
  \end{equation}
 As long as $m$ is sufficiently large, the term in the parentheses can be approximated by $N^2$ and the logarithmic correction can be neglected. The asymptotic late time regime sets in shortly after $\eta_0$ and the number density scales as the inverse volume.

 We observe that the number density depends entirely on the Early Universe mass parameter $m$, while the current physical particle mass plays essentially no role. This result is important for gravitational production of active and sterile neutrinos.
 The Early Universe mass is determined by the average value of the scalar field which enters the relevant Yukawa coupling. Since the de Sitter fluctuations drive the scalar fields to  values of order the Hubble rate or above, the parameter $m$ can be 
as large as $H$, making the  fermion production efficient.

 \subsection{Slow mass decrease}
 
 The fermion mass parameter $m(\eta)$ may also decrease slowly after inflation. This could be due to  thermal or non-perturbative effects in the postinflationary evolution. The resulting particle production efficiency changes compared to that of the sharp decrease case.

  \begin{figure}[h!]
    \centering
    \includegraphics[width=0.49\textwidth]{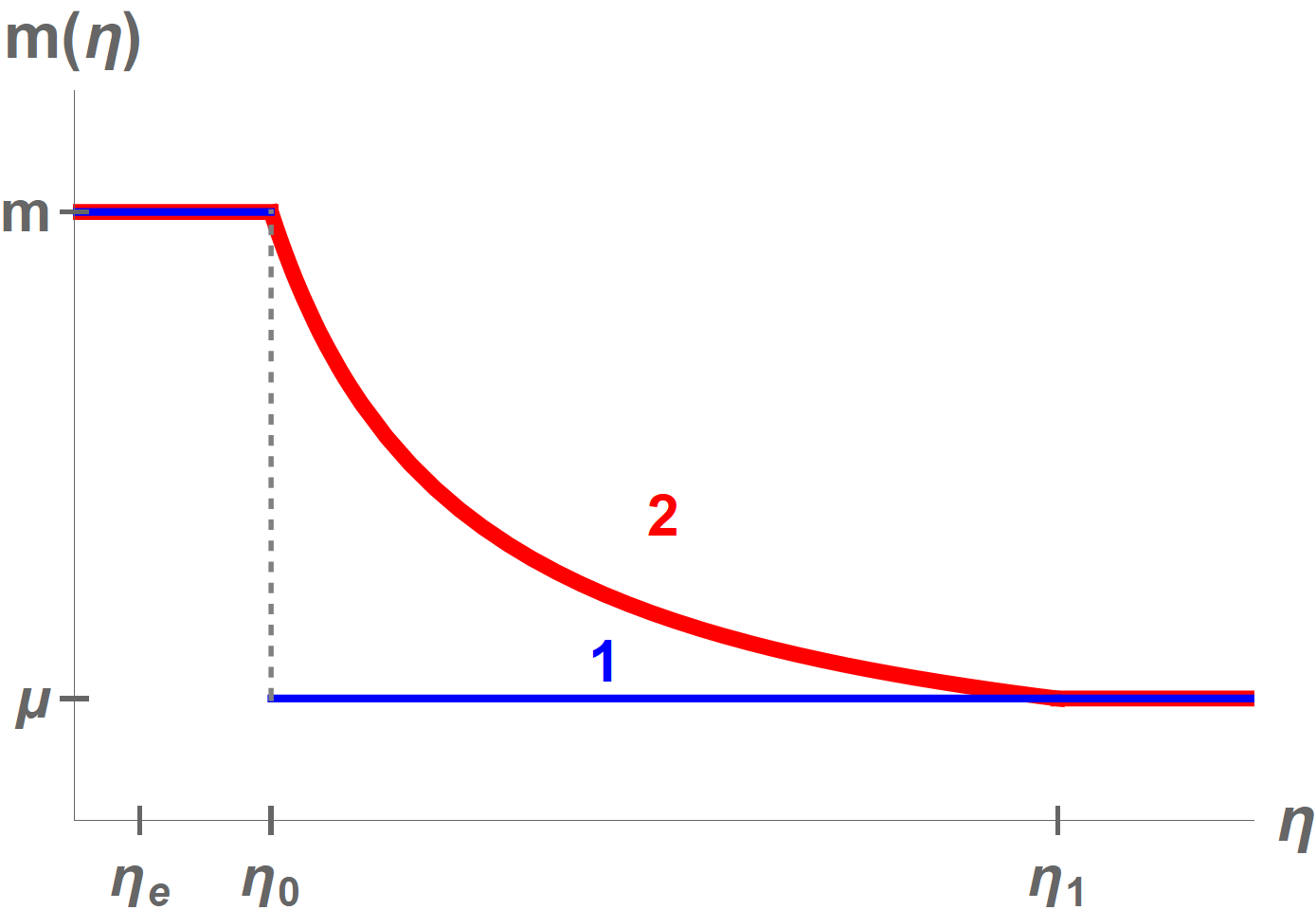}
     \includegraphics[width=0.49\textwidth]{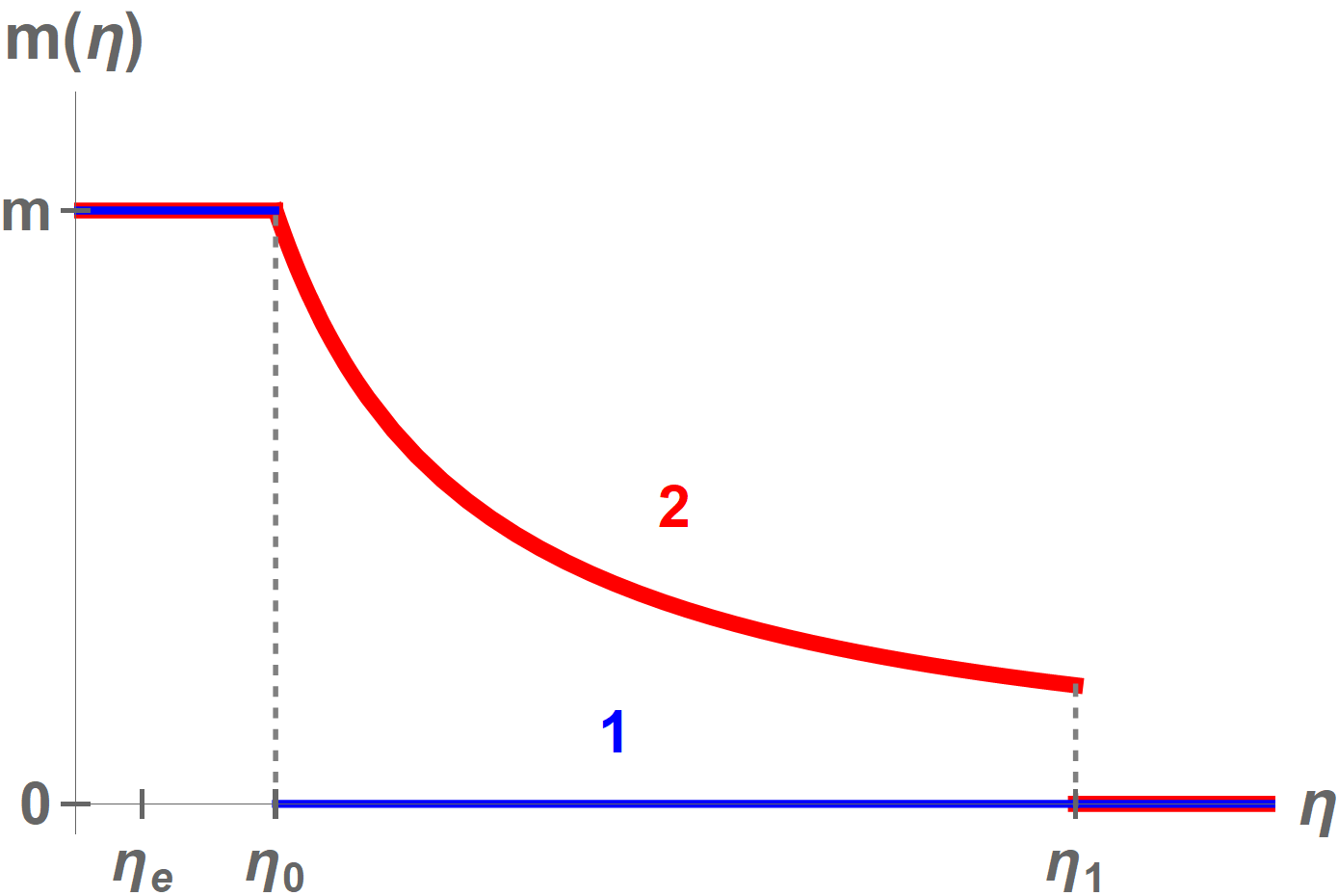}
    \caption{  {\small Time dependence of $m(\eta)$ in our numerical analysis (left) and analytical approximation (right). Curve 1: sharp mass decrease; curve 2: slow mass decrease.
    $\eta_e$ marks the end of inflation.}
    } 
    \label{m-profile}
\end{figure}

 Consider the following mass function (Fig.\,\ref{m-profile}, left):
 \begin{eqnarray}
 && \eta < \eta_0 ~: ~~m(\eta)=m={\rm const} ~, \nonumber\\
 && \eta_0 < \eta < \eta_1~: ~~ m(\eta) \propto 1/a ~,\nonumber\\
 && \eta_1 < \eta ~: ~~m(\eta)=\mu= {\rm const}  \;,
 \end{eqnarray}
where $\mu \ll M \ll m$ and $\eta_0> \eta_e$. We choose the function $m(\eta) \propto 1/a $ as motivated by the thermal mass contribution, which scales as $T \propto 1/a$.  
 
 This system can be solved numerically, while in our analytical estimates we may resort to further approximations.
 Let us take (Fig.\,\ref{m-profile}, right)
 \begin{equation}
 m(\eta< \eta_1) \gg M ~~, ~~ \mu\simeq 0\;.
 \end{equation}
 In this case, the early time dynamics is controlled entirely by $m(\eta)$, while at late times the fermion in question is effectively massless. Thus,
 $M$ does not enter $N_1(k)$, as before.

 The early and late time wavefunctions remain the same as in the previous section. This includes $\Phi (\eta_0)$ such that the only new ingredient is the $in$-wavefunction evolution between 
 $\eta_0$ and $\eta_1$. The Bogolyubov coefficient can now be computed at $\eta=\eta_1$.  
 
Consider the $in$-wavefunctions in the period $\eta_0 < \eta < \eta_1$. As before, the EOM's reduce to those of the single fermion case, see (\ref{EOM-comp-2}) and (\ref{uA-eq}).
 In regime $m(a) = m(a_0) \,a_0 /a$,  the EOM for all the $u$-components reads
\begin{equation}
u_j^{\prime\prime} + \left[a^2 m^2(a) +k^2\right]\, u_j=0\;,
\end{equation}
where $j= \{A_+,B_+, A_-,...\}$
For $k\ll a \,m(a)$, the oscillation frequency is 
\begin{equation}
\omega \simeq |a \, m(a)| =\,{\rm const }
\end{equation}
 and 
\begin{equation}
u_j (\eta ) = b_{1j} \, e^{i\omega \eta} + b_{2j} \, e^{-i \omega \eta} \;,
\end{equation}
with constant $b_{1j}, \, b_{2j}$.
The coefficients are fixed by the boundary conditions at $\eta_0$ in analogy with Eq.\,\ref{solutions}.
For example,  $u_{A_+}^{\rm in}$ at $\eta_0$ satisfies
 \begin{eqnarray}
 && u_{A_+}^{{\rm in\,}\prime} (\eta_0)   =  \left(-i a_0 m(a_0) \right)\,u_{A_+}^{\rm in}(\eta_0) \;.
 \end{eqnarray}
This relation is the same as for $e^{-i\omega \eta}$ with $\omega \simeq |a \, m(a)| $, so only one exponential survives in the sum and 
 \begin{equation}
u_{A_+}^{\rm in}(\eta)=  u_{A_+}^{\rm in}(\eta_0) \, e^{-i \omega (\eta- \eta_0)}\;.
\label{exp1}
\end{equation}
Similarly,
\begin{eqnarray}
&& u_{B_+}(\eta)=  u_{B_+}(\eta_0) \, e^{i \omega (\eta- \eta_0)}\;, \\
 && u_{A_{-(i)}}(\eta)=  u_{A_{-(i)}}(\eta_0) \, e^{i \omega (\eta- \eta_0)} \;\\
 &&  u_{B_{-(i)}}(\eta)=  u_{B_{-(i)}}(\eta_0) \, e^{-i \omega (\eta- \eta_0)}\;,
 \label{exp2}
 \end{eqnarray}
where $i=1,2$.
 The resulting ${\bf u}$-vectors are 
\begin{equation}
{\bf u}_{\rm in}^1 (\eta)\simeq {1\over \sqrt{2}} \,\left(
\begin{matrix}
\cos\omega (\eta-\bar \eta_0) \\
 \cos\omega (\eta-\bar \eta_0)  \\
- i\, \sin \omega (\eta-\bar \eta_0)  \\
i\, \sin \omega (\eta-\bar \eta_0)   
 \end{matrix}
\right)~~,~~
{\bf u}_{\rm in}^2 (\eta)\simeq    {1\over \sqrt{2}} \, \left(
\begin{matrix}
 - i\, \sin \omega (\eta-\bar \eta_0)    \\
i \,\sin \omega (\eta-\bar \eta_0)  \\
 \cos\omega (\eta-\bar \eta_0)  \\
\cos\omega (\eta-\bar \eta_0)   
 \end{matrix}
\right)  \;,
\label{app-eta0-1}
\end{equation}
with
\begin{equation}
   \bar \eta_0 = \eta_0 + {1\over \omega}\,   {m\over H_e} \, \left( - {N^2 \over 2} +    \ln {k\over a_e H_e}    \right) \;.  
   \end{equation}
The $out$ vector ${\bf v}^1_{\rm out}$ remains the same as before, i.e. (\ref{out-v-0}).

\subsubsection{Particle number }

We can compute the Bogolyubov coefficient at $\eta=\eta_1 \gg \bar \eta_0$.
 For momenta below the critical value 
 \begin{equation}
k < \tilde k_* = a_0\, m(a_0) \;,
\end{equation}
 the  occupation number is 
 \begin{equation}
N_1(k) = \sin^2 \omega\eta_1\;.
\end{equation}
This value oscillates, however, the moment $\eta_1$ cannot be defined on a scale smaller than the inverse effective particle mass, i.e. $1/\omega$. Averaging over the oscillation period,
we obtain the average occupation number
 \begin{equation}
\langle N_1 (k) \rangle = 1/2 \;.
\end{equation}

 \begin{figure}[h!]
    \centering
    \includegraphics[width=0.49\textwidth]{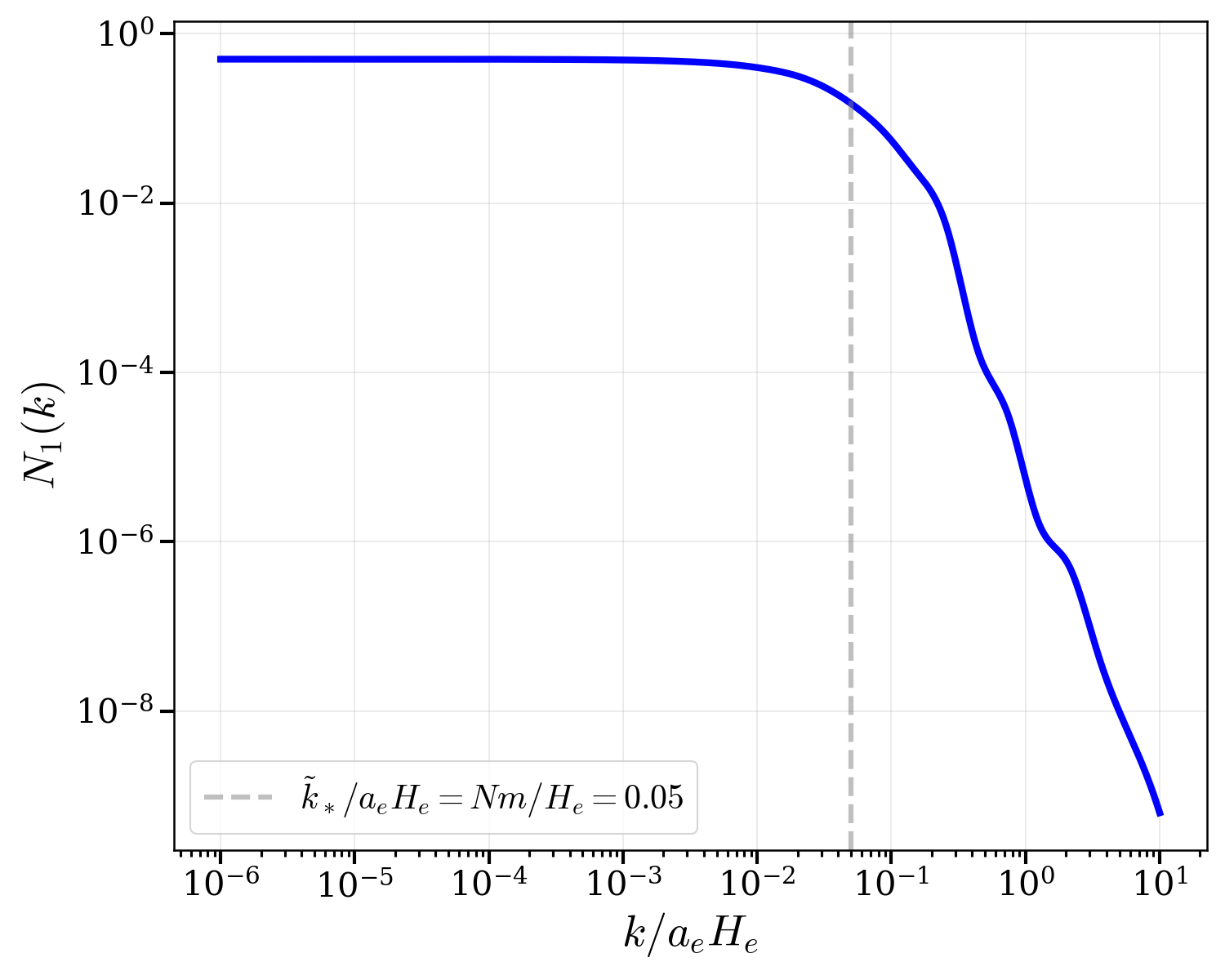}
    \caption{ The average occupation  number $N_1(k)$ for a slowly decreasing $m(\eta)$, with
    $m/H_e = 10^{-2}, M/H_e = 10^{-5}, N = 5, \mu \simeq 0 $. The dashed line marks the cutoff $\tilde k_*= a_0 m(a_0)$, beyond which particle production drops. 
    } 
    \label{N1-slow}
\end{figure}

 For momenta far above $\tilde k_*$, the frequency is dominated by $k$: $\omega \simeq k$. Since $a_0 m(a_0)/\omega \rightarrow 0$, the boundary conditions at $\eta_0$ enforce
 the universal wavefunction form proportional to $\cos\omega (\eta - \eta_0)$ instead of the exponential in (\ref{exp1}) and (\ref{exp2}). This results in a sharp  $N_1(k)$ decrease such that
 $\tilde k_*$ serves as a momentum cutoff.

 We note that,  
   in the intermediate period    $\eta_0 < \eta < \eta_1$, the Hamiltonian is time-independent and particle production should not occur. 
   As seen above, the particle number on the average
 remains constant  and independent of $\eta_1$. 
 This implies that  all the particle production occurs close to $\eta_0$, when the EOM regime changes.

  Our numerical results are shown in Fig.\,\ref{N1-slow}. Again, we find good agreement with our analytical estimates. The non-zero neutrino mass corrections scale as  $\mu/M,\, \mu/H_e$ and thus are irrelevant.

 The total physical number density of the produced light fermions is obtained via (\ref{n0}) with the replacement $k_* \rightarrow \tilde k_*$, yielding
 \begin{equation}
 n_1 \simeq {\# {\rm d.o.f.}} \times {N^3 m^3\over 12 \pi^2  }   \, \left( {a_e\over a}\right)^3 \;.
 \label{n1-tot2}
  \end{equation}
 This result is enhanced by the factor $\left( N \sqrt{m/H_e}\,\right)^{-1} \gg 1$ compared to the sharp mass decrease case. As before, the current physical fermion mass plays essentially no role.

 \section{Discussion}

We have studied gravitational fermion production     with a seesaw-like mass matrix.
Our main finding is that particles which are very light or even massless  at present, can be very efficiently produced by gravity in the Early Universe. This is due to the large off-diagonal entries in the mass matrix induced by the scalar
field condensate, which takes on a value of the order of   the Hubble rate  in the Early Universe. 
 In terms of the seesaw-like mass matrix
 (\ref{m-matrix}), the production regime is controlled by the mass matrix at $\eta \sim 0$, while the current physical mass is determined by its behavior at $\eta \rightarrow \infty$,
 \begin{eqnarray}
&&{\rm production~stage:}~~m(0) \nonumber\\
&&  {\rm present~stage:}~~{m(\infty)^2\over M}        \lll m(0)               \nonumber
\end{eqnarray} 
 The latter can be vanishingly  small, yet the particle abundance is determined entirely by $m(\eta)$ in the Early Universe.

 These results apply directly to the active neutrinos, up to the number of degrees of freedom included in $n_1$. The Dirac mass is given by $m(\eta)= Y_\nu \,v (\eta)/\sqrt{2}$, and    can be larger than the Majorana mass $M$ during inflation. Its value   controls the neutrino production in the Early Universe
with  the maximum being achieved at $m(\eta)\lesssim H_e \lesssim 10^{13}\,$GeV, where the bound on the Hubble rate is imposed by the inflationary data \cite{Planck:2018jri,BICEP:2021xfz}.  Thus, for our estimate (\ref{n1-tot2}) to be applicable, the Majorana mass is bounded by $M\ll 10^{13}\,$GeV. 
 Currently, however, the active neutrino masses are determined by the seesaw relation $m_\nu \sim m(\infty)^2/M$ and constrained to be very small by cosmology \cite{DESI:2024mwx}. Despite that, their gravitational production efficiency can be on par with that of the top quark.
 Yet,  in standard reheating models, this plays no role since the active neutrinos thermalize and their  abundance is controlled by the temperature (see \cite{Dolgov:2002wy}  for a review).

 Our considerations also apply to the system of light and heavy right-handed neutrinos $\nu_{R_a}$. This possibility is more interesting since these fermions may interact very weakly and never thermalize. Their Majorana mass matrix is determined by the Yukawa-type interactions
 with singlet scalars $S_i$, i.e. $S_i \nu_{R_a} \nu_{R_b}$,     and 
  can have the form (\ref{m-matrix})
 due to symmetries. 
Owing  to the mixing term, the production of the lighter state may be very efficient. The latter, if long-lived,  can  
  contribute to the energy density of dark matter. 
 This possibility will be studied in detail in our subsequent work. We note that gravitationally produced $\nu_R$ create a non-thermal neutrino background of the type discussed in   \cite{Chen:2015dka}, although their number density is typically far below that assumed in 
 this reference.

 \begin{figure}[h!]
    \centering
    \includegraphics[width=0.49\textwidth]{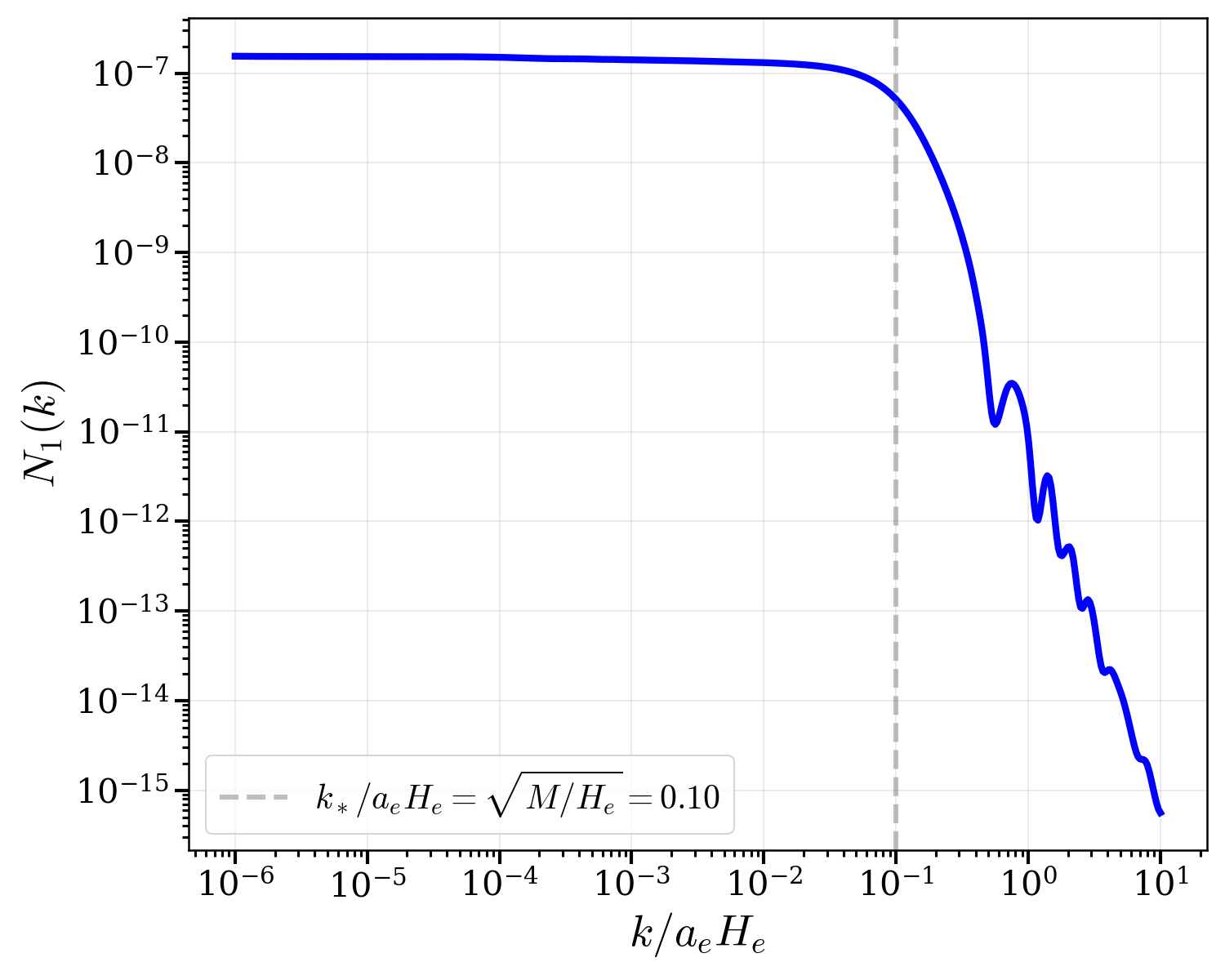}
    \caption{The average occupation  number $N_1(k)$  in the small mixing regime $m\ll M$, with a slowly decreasing 
     $m(\eta)$. The parameters are 
    $m/H_e = 10^{-5}, M/H_e = 10^{-2}, N = 5, \mu \simeq 0 $ and   $k_*(M)= a_e\,\sqrt{MH_e}$.
    } 
    \label{N1-small-m}
\end{figure}

 The particle density results (\ref{n1-tot1}) and \ref{n1-tot2}) are consistent with the corresponding single fermion formulae  in Ref.\,\cite{Feiteira:2025phi}. Therefore, our system in the {\it maximal mixing} regime $m(0) \gg M\;,\; m(\infty)\sim 0$
   is  similar  to a single fermion with decreasing in time mass.
 The difference is that, in the current case, the late time mass can be almost arbitrarily small, while, in the single fermion case, it  is fixed by the Yukawa coupling and the late-time scalar VEV. 
 Furthermore, the 2 fermion system allows for more complicated evolution patterns, e.g. when $m(\eta) $ crosses the value of $M$ after inflation.
 In the regime $m(0) \ll M$, the light state production is suppressed by the mixing angle $\sin\phi \sim m(0)/M$. Fig.\,\ref{N1-small-m} shows the high-$k$ end of the spectrum, which is cut off at $k_*(M) =a_e \sqrt{MH_e}$ characteristic of  heavy fermion production with constant 
 mass.\footnote{The light fermion production channel via the mixing gives the main contribution to the particle density $n_1$.}
 The corresponding occupation number is determined by the light $out$ fermion component in the $in$ heavy mass eigenstate, i.e. 
  $N_1 \sim 1/2 \times \left(m(0)/M\right)^2$.

 The particle abundance  $Y$ for the fast and slow mass decrease corresponding to the number densities (\ref{n1-tot1}) and \ref{n1-tot2}) is given by
\begin{equation}
Y^{\rm fast} \sim 10^{-3}\times N^4 \; {m^{7/2} \over H_e^2 \, M^{3/2}_{\rm Pl} }~~~,~~~
Y^{\rm slow} \sim 10^{-3}\times N^3 \; {m^{3} \over \left(H_e \, M^{}_{\rm Pl}\right)^{3/2} }\;.
\end{equation}
 Here the abundance is defined in the usual way $Y\equiv n/ s_{\rm SM}$, where $s_{\rm SM}$ is the SM entropy density, and we have assumed radiation domination after inflation. 
 The parameters $m$ and $N$ characterize  the effective fermion mass during inflation and the lifetime of the scalar condensate after inflation, respectively. These are model-dependent and, while for the Higgs field they are 
 well understood, the  singlet scalar condensate that determines the Majorana mass can have very different properties.

 The maximal abundance is achieved in the limit $N \sqrt{m/H_e} \sim 1$ \cite{Feiteira:2025phi}, which corresponds to a heavy fermion whose mass is effectively constant within the entire production period ($H(\eta) \gtrsim m$).
 The gravitational production is efficient only if $m\lesssim  H_e\lesssim 10^{13}\,{\rm GeV}$, which sets the absolute upper bound
 \begin{equation}
 Y \lesssim 10^{-3}\times \left(  {m\over M_{\rm Pl}}    \right)^{3/2} \lesssim 10^{-11} \;,
 \label{Y-bound}
 \end{equation}
 similarly to the single-fermion case \cite{Feiteira:2025phi}.
 For comparison,
 the abundance of dark matter
 is $Y_{\rm DM}=4\times 10^{-10} \, {{\rm GeV}/m_{\rm DM}}$. Therefore, our gravitational production mechanism can generate the right amount of dark matter only if $m_{\rm DM} \gtrsim {\cal O}(10)\,{\rm GeV}$.
 This bound is not altered if radiation domination is replaced with matter domination since in the latter case gravitational production is less efficient \cite{Feiteira:2025phi}.
 We note that the particle number is approximately conserved only when the particle couplings are small enough, e.g. the fermions do not thermalize, and the above considerations are meaningful under this assumption.

 An additional gravitational fermion production source is provided by the inflaton-fermion coupling induced by classical gravity \cite{Ema:2015dka}. If the inflaton field    $\phi$  oscillates after the end of inflation, this 
 creates a time-dependent background entailing 
  fermion production. The corresponding coupling
 has the form ${\cal O}(10^{-1})\times {m\over M_{\rm Pl}^2} \phi^2 \bar \Psi \Psi$ and the resulting particle abundance has been computed in \cite{Koutroulis:2023fgp}.  The latter is maximized in the radiation-like background corresponding to the (locally) quartic inflaton potential
 ${1\over 4} \lambda \phi^4$.
 For a fermion with effective mass $H_e$, the production occurs for a brief period of time, when the effective inflaton mass is large enough, after which the process gets suppressed kinematically. We find that the resulting particle abundance is somewhat smaller than the 
 above maximal value, typically on the order of $10^{-12}$. Thus, this process can be treated as subleading. Note also that  the bound (\ref{Y-bound}) applies more generally, even if the inflaton field does not undergo oscillations around the potential minimum.

 The quantum gravity induced operator   ${{\cal C}\over M_{\rm Pl}} \phi^2 \bar \Psi \Psi$ can, on the other hand, be very efficient in particle production \cite{Koutroulis:2023fgp,Lebedev:2022ljz}. Its Wilson coefficient ${\cal C}$ is essentially unconstrained and can only be computed in a UV complete quantum 
 gravity theory. The analysis of Ref.\,\cite{Koutroulis:2023fgp} shows that even a small ${\cal C}$ is sufficient to generate all the dark matter in the Universe. In our current work, we restrict ourselves to well-understood  classical gravitational effects and have nothing to add about quantum gravity.

  \section{Conclusion}
  
  We have developed the Bogolyubov coefficient formalism for gravitational production of fermions with time-dependent mixing. This allows us to evaluate inflationary production of active and sterile neutrinos, whose masses and mixings
  are environment-dependent in the Early Universe. In particular, the Dirac  neutrino masses depend on the Higgs field condensate, which can reach very  large  values due to de Sitter fluctuations. This makes gravitational neutrino 
  production much more efficient, even if the neutrinos  are   nearly  massless at late times.  Similarly, light sterile neutrino production is controlled by the scalar-dependent Majorana  mass and can also be enhanced. This phenomenon may have interesting implications for dark matter.
  Finally, we find that the abundance of active and sterile neutrinos produced by classical gravity is bounded by $Y \lesssim 10^{-11}$. This bound, however, may be violated in quantum gravity.
  \\ \ \\
  {\bf Acknowledgements.} This work was supported by the National Research Foundation of Korea grants funded by the Ministry of Science and ICT (RS-2024-00356960 and RS-2025-00559197) and by the National Supercomputing Center with supercomputing resources including technical support (KSC-2025-CRE-0282).

\end{document}